\documentclass[12pt]{article}

\usepackage{amsmath, amssymb, slashed}
\usepackage[dvips]{graphicx}
\usepackage{epsf}

\begin{document}

\begin{titlepage}
\begin{center}

\hfill UT-HET-063 \\
\hfill IPMU12-0029 \\
\hfill HGU-CAP-14 \\

\vspace{1.5cm}
{\Large\bf Vector WIMP Miracle}
\vspace{1.5cm}

{\bf Tomohiro Abe}$^{(a)}$,
{\bf Mitsuru Kakizaki}$^{(b)}$, \\
{\bf Shigeki Matsumoto}$^{(c)}$,
and
{\bf Osamu Seto}$^{(d)}$

\vspace{1.0cm}
{\it
$^{(a)}${\it Institute of Modern Physics and Center for High Energy Physics, 
Tsinghua University, Beijing 100084, China} \\ 
$^{(b)}${\it Department of Physics, University of Toyama, Toyama, Japan} \\
$^{(c)}${\it IPMU, TODIAS, University of Tokyo, Kashiwa, Japan} \\
$^{(d)}${\it Department of Architecture and Building Engineering, \\
Hokkai-Gakuen University, Sapporo, Japan} \\
}
\vspace{2cm}

\abstract{
Weakly interacting massive particle (WIMP) is well known to be a good candidate for dark matter, and it is also predicted by many new physics models beyond the standard model at the TeV scale. We found that, if the WIMP is a vector particle (spin one particle) which is associated with some gauge symmetry broken at the TeV scale, the higgs mass is often predicted to be 120--125 GeV, which is very consistent with the result of higgs searches recently reported by ATLAS and CMS collaborations at the Large Hadron Collider experiment. In this letter, we consider the vector WIMP using a non-linear sigma model in order to confirm this result as general as possible in a bottom-up approach. Near-future prospects to detect the vector WIMP at both direct and indirect detection experiments of dark matter are also discussed.}

\end{center}
\end{titlepage}
\setcounter{footnote}{0}

%%%%%%%%%%%%%%%%%%%%%%%%
%%%%% Introduction %%%%%
%%%%%%%%%%%%%%%%%%%%%%%%
\section{Introduction}
\label{sec: intro}

There are many compelling evidences for the existence of dark matter in our universe, and many experimental efforts are presently devoted to detect the dark matter directly and indirectly~\cite{Bertone:2010zz}. On the other hand, because the detailed nature of the dark matter is not revealed yet, many dark matter candidates have been proposed so far from the viewpoint of new physics beyond the standard model (SM) at the TeV scale. Among those, the weakly interacting massive particle (WIMP), whose mass is postulated to be 10--1000 GeV, is known to be a good candidate for dark matter, because it can easily satisfy all constraints imposed by cosmological and astrophysical dark matter experiments and naturally explain the dark matter abundance observed today~\cite{Komatsu:2010fb}. In this letter, we especially focus on the vector (spin one) WIMP which is associated with some gauge symmetry broken at the TeV scale. Since the vector particle acquires its mass from the symmetry breaking, the mass is predicted to be 100--1000 GeV, which is very consistent with the WIMP hypothesis.

The simplest model of the vector WIMP dark matter is described based on
SU(2)$_L$ $\times$ U(1)$_1$ $\times$ U(1)$_2$ gauge symmetry. In order
to guarantee the stability of the dark matter, the Z$_2$ symmetry is
also imposed by postulating that the lagrangian of the model is
invariant under the exchange of U(1)$_1$ and U(1)$_2$ gauge
interactions. The U(1)$_1$ $\times$ U(1)$_2$ symmetry is assumed to be
broken at the TeV scale into the diagonal U(1), which is identified with
the SM gauge interaction of U(1)$_Y$. This fact means that the dark
matter particle is provided as the partner of the hyper-charge gauge
boson. As will be discussed in section \ref{sec: model}, the strength of
the coupling between two higgs bosons and two dark matter particles is
definitely given by $(g^{\prime})^2/4$ with $g^\prime$ being the
U(1)$_Y$ gauge coupling.\footnote{Opposite directions that reconcile the
WMAP region with the LHC excess by adjusting the interactions between the
Higgs boson and vector dark matter have been discussed in Higgs-portal dark
matter models \cite{OtherVectorWIMPs}.}
%On the contrary, adjustable
%interactions between the Higgs boson and dark matter have been discussed
%in Higgs-portal dark matter
%models\cite{OtherVectorWIMPs}.} 
This simplest model is  
embedded in several realistic models for the new physics predicting the
vector WIMP such as universal extra-dimension models~\cite{UED} and
little higgs models with T-parity~\cite{LH}.

When no other new particles, which could be predicted in the new physics at the TeV scale, are degenerated in mass to the vector WIMP dark matter, the annihilation of the dark matter is governed by the process into $W(Z)$ boson pair through the s-channel exchange of the higgs boson. Its annihilation cross section and, as a result, the thermal relic abundance of the dark matter therefore depend only on the masses of dark matter and higgs boson. The abundance turns out to be consistent with the WMAP observation when the higgs mass is within the range of 120--125 GeV as will be shown in section \ref{sec: results}, which is very attractive from the viewpoint of the result of higgs searches recently reported by ATLAS and CMS collaborations at the Large Hadron Collider (LHC) experiment~\cite{LHC H}. Interestingly, this result is insensitive to the dark matter mass ($m_{\rm DM}$) as long as $m_{\rm DM} \sim 100$ GeV. According to this result, we also discuss future prospects to discover the vector WIMP in direct and indirect detection experiments of dark matter in this section. We found that the signal of the dark matter can be discovered at both experiments in the near future.

%%%%%%%%%%%%%%%%%%%%%%%%%%
%%%%% Simplest model %%%%%
%%%%%%%%%%%%%%%%%%%%%%%%%%
\section{Simplest model for the vector WIMP}
\label{sec: model}

\begin{figure}[t]
\begin{center}
\includegraphics[scale=0.7]{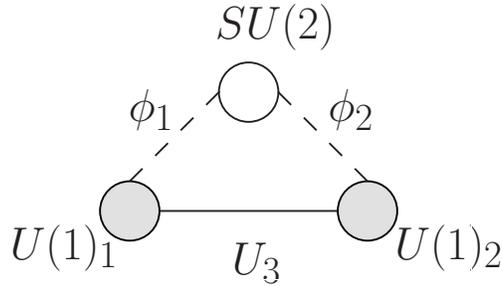}
\caption{\small The structure of the SU(2)$_L$ $\times$ U(1)$_1$ $\times$ U(1)$_2$ model for the vector WIMP dark matter expressed by using the moose notation. See the text for more details.}
\label{fig: moose}
\end{center}
\end{figure} 

We consider the simplest model for the vector WIMP dark matter in this
 section, which is described based on the SU(2)$_L$ $\times$ U(1)$_1$
 $\times$ U(1)$_2$ gauge theory in its electroweak sector. The structure
 of the gauge-higgs sector in this model involving symmetries and their
 breaking patterns is schematically expressed by using the `moose'
 notation~\cite{Georgi:1985hf} as shown in Fig.\ref{fig: moose}, where
 the white circle stands for the SM SU(2)$_L$ gauge symmetry, while
 black ones are U(1) gauge symmetries (U(1)$_1$ and U(1)$_2$). On the
 other hand, the solid line represents the non-linear sigma field, $U_3
 \equiv \exp (i \pi_3/v_3)$, and broken lines are linear sigma fields,
 namely, higgs fields denoted by $\phi_1$ and $\phi_2$. All of
 non-linear sigma and higgs fields spontaneously break the symmetries
 connected to them into their diagonal ones. In order to guarantee the
 stability of the vector WIMP, the Z$_2$ symmetry is imposed by
 postulating that the model is invariant under the exchange of U(1)$_1$
 and U(1)$_2$ gauge interactions, which can be expressed by the symmetry
 under the left-right reflection of the diagram in Fig.\ref{fig: moose}:
%namely
$\phi_1 \to \phi_2$,
$\phi_2 \to \phi_1$,
and
$U_3 \to U_3^{*}$.
 Both gauge couplings of U(1)$_1$ and U(1)$_2$ interactions as well as vacuum expectation values (VEVs) of two higgs fields $\phi_1$ and $\phi_2$ are therefore taking the same value. \\

With the use of the diagram shown in Fig.\ref{fig: moose}, the lagrangian (kinetic terms) of non-linear sigma and higgs fields ($U_3$, $\phi_1$, and $\phi_2$) are given as follows;
\begin{eqnarray}
{\cal L}_{\rm higgs}
=
\left( D_\mu \phi_1^\dagger D^\mu \phi_1 \right)
+ \left( D_\mu \phi_2^\dagger D^\mu \phi_2 \right)
+ \frac{v_3^2}{2} \left( D_\mu U_3^\dagger D^\mu U_3 \right)
- V(\phi_1, \phi_2, U_3),
\label{eq: higgs_kin}
\end{eqnarray}
where the higgs field $\phi_i$ ($i = 1, 2$) is decomposed to be $\phi_i =  (v_i + h_i + i \tau^a \pi_i^a) \cdot (0, 1)^T/\sqrt{2}$ with $\tau^a$ and $v_i$ being the Pauli matrix and the VEV of the higgs field, respectively. The higgs potential is simply denoted by $V(\phi_1, \phi_2, U_3)$. The covariant derivatives of the non-linear sigma field and higgs fields are defined by following equations,
\begin{eqnarray}
D_{\mu} \phi_i
&=&
\partial_{\mu} \phi_i
+ i g (\tau^a/2) W^a_\mu \phi_i
+ i (g'_i/2) B^{(i)}_\mu \phi_i, \\
D_\mu U_3
&=&	
\partial_\mu U_3
+ i (Y_3/4) g'_1 B^{(1)}_\mu U_3
- i (Y_3/4) g'_2 B^{(2)}_\mu U_3,
\end{eqnarray}
where $W^a_\mu$ and $B_\mu^{(i)}$ are SU(2)$_L$ and U(1)$_i$ gauge
 fields, respectively. Because of the Z$_2$ symmetry, we take $g'_1 =
 g'_2$ as well as $v_1 = v_2$ later, where the gauge coupling $g'_1 =
 g'_2$ relates to that of the SM U(1)$_Y$.
The U(1)$_1$ and U(1)$_2$ charges of the non-linear
sigma field $U_3$ are set to be $Y_3/4$ and $-Y_3/4$, respectively,
in order to have the Z$_2$ symmetry, which makes
$\pi_3$ an Z$_2$ odd particle (would-be NG boson).

As can be seen from the lagrangian, the model has five gauge bosons and seven NG bosons. Four NG bosons are eaten and the gauge bosons except photon become massive. The rest of three NG bosons become pseudo-NG bosons, because we can write down gauge invariant mass terms such as $(\phi_1^\dagger \phi_2)(U_3)^{2/Y_3}$ which is naturally involved in the higgs potential $V(\phi_1, \phi_2, U_3)$. There are five particles which are odd under the Z$_2$ parity; a linear combinations of the $U(1)$ gauge fields $(B^{(1)}-B^{(2)})/\sqrt{2}$ which is nothing but the vector WIMP, a linear combination of the higgs fields $(h_1-h_2)/\sqrt{2}$, and three linear combinations of the pseudo-NG bosons. The later four scalars acquire their masses through the higgs potential, so that they can be heavy enough compared to the vector WIMP. We therefore concentrate on the vector WIMP and other particles which are even under the Z$_2$ symmetry in following discussions. Mass matrices of the gauge bosons are eventually summarized as follows,
\begin{eqnarray}
\frac{g^2 v^2}{4} W^+_\mu W^{-\mu}
+ \frac{v^2}{8}
\begin{pmatrix} W^3_\mu & B_\mu \\ \end{pmatrix}
\begin{pmatrix}
g^2 & - g g' \\
- g g' & g^{\prime 2} \\
\end{pmatrix}
\begin{pmatrix} W^{3 \mu} \\ B^\mu \end{pmatrix}
+ \frac{g^{\prime 2}}{8} \left(v^2 + Y_3^2v_3^2 \right) V_\mu V^\mu,
\end{eqnarray}
where $v \equiv \sqrt{2} v_1 = \sqrt{2} v_2$ and $g' \equiv g'_1/\sqrt{2} = g'_2/\sqrt{2}$, while $W_\mu^\pm = (W_\mu^1 \mp i W_\mu^2)/\sqrt{2}$, $B_\mu \equiv (B^{(1)}_\mu + B^{(2)}_\mu)/\sqrt{2}$, and $V_\mu \equiv (B^{(1)}_\mu - B^{(2)}_\mu)/\sqrt{2}$. With the use of the Weinberg angle defined by $\sin\theta_W \equiv s_W^2 = g^{\prime 2}/(g^2 + g^{\prime 2})$ ($\cos\theta_W = c_W^2 = 1 - s_W^2$) and the $Z$ boson mass $m_Z$, the mass of the vector WIMP dark matter is expressed by
\begin{eqnarray}
m_{\rm DM}
\equiv m_{Z} s_W \sqrt{1 + Y_3^2v_3^2/v^2}
\simeq 178~{\rm GeV} (Y_2 v_3/1~{\rm TeV}).
\end{eqnarray}
It can be seen that the mass of the vector WIMP dark matter is {\cal O}(100) GeV when the breaking scale associated with U(1)$_1$ $\times$ U(1)$_2$ $\to$ U(1)$_Y$ is the TeV scale.

%Before going to close the section,
 We consider interactions between higgs boson and vector WIMP dark matter, which are the most important interactions to discuss phenomenology of the dark matter, as will be seen in the next section. Kinetic terms of the higgs bosons $\phi_1$ and $\phi_2$ involve following gauge interactions,
\begin{eqnarray}
\left[
\frac{1}{2} \left( v_1 + h_1 \right)^2
+ \frac{1}{2} \left( v_1 + h_2 \right)^2
\right]
\frac{g^{\prime 2}}{4}V_\mu V^\mu
=
\frac{g^{\prime 2}}{8} V_\mu V^\mu h^2
+
\frac{g^{\prime 2} v}{4} V_\mu V^\mu h
+
\cdots,
\end{eqnarray}
where $h \equiv (h_1 + h_2)/\sqrt{2}$. This scalar particle is identified with the SM-like higgs boson, because it is the only Z$_2$-even scalar which remains as a physical state. It is very important to notice that the interactions between higgs boson and vector WIMP are governed by the gauge coupling of the SM U(1)$_Y$ interaction.

% --------------------------------------------------
% Added for the comment 1 from the refree
% --------------------------------------------------

Before closing this section, we mention the fermion sector for the
sake of completeness although the structure of fermions does not
affect the properties of the vector WIMP dark matter, which are
discussed in the subsequent sections.  
%For each generation the
%SU(2)$_L$ $\times$ U(1)$_1$ $\times$ U(1)$_2$ quantum numbers of the
%left-handed quark $q_L$, right-handed up-type quark $u_R$ and
%right-handed down-type quark $d_R$ are assigned as
%%
%\begin{eqnarray}
%  \begin{array}{c||c|c|c}
%    & {\rm SU(2)}_L &{\rm U(1)}_1 &{\rm U(1)}_2\\ \hline
%    q_{L} & {\bf 2} &1/12 &1/12\\
%    u_{R} & {\bf 1} &1/3 &1/3\\
%    d_{R} & {\bf 1} &-1/6 &-1/6\\
%  \end{array}
%\end{eqnarray}
The assignment of the 
SU(2)$_L$ $\times$ U(1)$_1$ $\times$ U(1)$_2$ quantum numbers for the
left-handed quark $q_L$, right-handed up-type quark $u_R$ and
right-handed down-type quark $d_R$
is given in table \ref{tab:fermion}.
\begin{table}
\begin{center}
   \begin{tabular}{c||c|c|c}
    & ${\rm SU(2)}_L$ & ${\rm U(1)}_1$ & ${\rm U(1)}_2$\\ \hline
    $q_{L}$ & {\bf 2} &1/12 &1/12\\
    $u_{R}$ & {\bf 1} &1/3 &1/3\\
    $d_{R}$ & {\bf 1} &-1/6 &-1/6\\
\end{tabular}
\end{center}
\caption{Quantum numbers for the quark sector.}
\label{tab:fermion}
%\ref{tab:fermion}
\end{table}
It is clear from this expression that the quark fields are $Z_2$-invariant.
The SM quark mass terms stem from the following $Z_2$-invariant
Yukawa interactions: 
\begin{eqnarray}
{\cal L}_Y & = & -
y_u
\overline{q}_{L} 
\tilde{\phi}_1
U_3^{1/Y_3}
u_R
-
y_u
\overline{q}_{L} 
\tilde{\phi}_2
(U_3^{*})^{1/Y_3}
u_R
\nonumber \\
&& -
y_d
\overline{q}_{L} 
\phi_1
(U_3^{*})^{1/Y_3}
d_R
-
y_d
\overline{q}_{L} 
\phi_2
U_3^{1/Y_3}
d_R
+
({\rm h.c.}),
\end{eqnarray}
where $\tilde{\phi}_i = i \tau^2 \phi_i^{*}$.  Here, $y_u$ and $y_d$
denote the up- and down-type Yukawa coupling constants, respectively,
and generation indices are omitted for simplicity.  The same
discussion is applicable to the lepton sector.

%%%%%%%%%%%%%%%%%%%%%%%%%%%%
%%%%% DM phenomenology %%%%%
%%%%%%%%%%%%%%%%%%%%%%%%%%%%
\section{Vector WIMP predictions}
\label{sec: results}

We are now at the position to discuss what kinds of phenomenological consequences we have by considering the vector WIMP dark matter. The prediction of the vector WIMP we first discuss about is on the mass of the higgs boson, which is deduced from the thermal relic abundance of the dark matter. After that, we discuss several signals of the vector WIMP at both direct and indirect detection experiments.

\subsection{Thermal relic abundance}

The thermal relic abundance of the vector WIMP dark matter is obtained numerically by integrating the following Boltzmann equation which describes the number density of the dark matter particle (denoted by $n$) in the early universe~\cite{Boltzmann},
\begin{equation}
\frac{dn}{dt} + 3Hn
=
-\langle \sigma v \rangle (n^2 - n_{\rm EQ}^2),
\end{equation}
where $H$, $\langle\sigma v\rangle$, and $n_{\rm EQ}$ are the Hubble parameter, the annihilation cross section (times relative velocity) of the vector WIMP which is thermally averaged, and its number density in thermal equilibrium, respectively. Since the vector WIMP annihilates into $W$ and $Z$ boson pairs through the s-channel exchange of the higgs boson and the $h-V-V$ vertex is fixed by $g^\prime$, the annihilation cross section depends only on the masses of vector WIMP and higgs boson. As a result, the thermal relic abundance of the dark matter also depends only on these two parameters.

\begin{figure}[t]
\begin{center}
\includegraphics[scale=0.6]{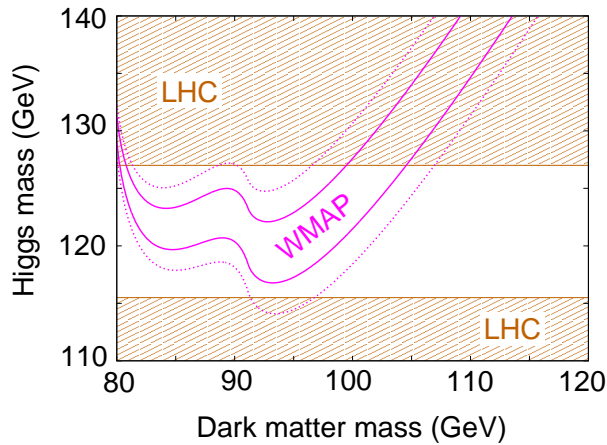}
\caption{\small Parameter region consistent with the WMAP observation at 68\% (solid line) and 95\% (dotted line) C.L. on the ($m_{\rm DM}$, $m_h$)-plane. The regions which are constrained by current higgs searches at the LHC experiment are also shown as shaded ones.}
\label{fig: omegah2}
\end{center}
\end{figure}

Numerical result of the abundance is shown in Fig.\ref{fig: omegah2}, where the parameter region which is consistent with the WMAP observation at 68\% (solid line) and 95\% (dotted line) C.L. are depicted on the ($m_{\rm DM}$, $m_h$)-plane. In order to calculate the abundance, we have used {\tt micrOMEGAs}~\cite{MicrOMEGAs} after implementing our model into the code. Appropriate modifications are made by using {\tt LanHEP}~\cite{LanHEP}.  We have also shown regions which are constrained by current higgs searches at the LHC experiment as shaded ones. It can be seen from the figure that the higgs mass is predicted to be 120--125 GeV, which is very consistent with that strongly suggested by the higgs searches. It is also worth noting that, if the vector WIMP mass is less than 100 GeV, the abundance is not sensitive to the mass and the higgs mass favored by the WMAP observation keeps staying around $m_h \simeq$ 120--125 GeV.

%-------------------------------------------------- 
% added in version 2.
%-------------------------------------------------- 
Around $m_{DM} = m_h/2$, there are also regions consistent with the
WMAP and LHC bounds.  Since these Higgs-pole regions require
fine-tuning between the mass parameters, they are disregarded in this
manuscript.

\subsection{Direct detection}

We next consider the signal of the vector WIMP at direct detection experiments of dark matter. The scattering between the dark matter and a nucleon (proton) occurs by exchanging the higgs boson. Since the $h-V-V$ vertex is fixed by the U(1)$_Y$ gauge coupling, the cross section depends only on $m_{\rm DM}$ and $m_h$ as in the case of the relic abundance. The scattering cross section is obtained by the formula~\cite{Direct};
\begin{equation}
\sigma_{\rm SI}^{(p)}
=
\frac{1}{\pi}
\left( \frac{m_p}{m_p + m_{\rm DM} }\right)^2 f_p^2, 
\end{equation}
where $m_p$ is the proton mass and the coefficient $f_p$ is the (spin-independent) coupling between the dark matter and a proton which is phenomenologically evaluated by 
\begin{equation}
\frac{f_p}{m_p}
=
\sum_{q = u, d, s} f_{Tq}^{(p)} \frac{\alpha_q}{m_q} 
+ \frac{2}{27} f_{TG}^{(p)} \sum_{q = c, b, t} \frac{\alpha_q}{m_q}.
\end{equation}
Here, $m_q$ is a quark mass and $\alpha_q$ is the coupling in front of the effective interaction, $\alpha_q V^\mu V_\mu \bar{q} q$, which is obtained by integrating the higgs field out from the original lagrangian. The coefficients $f_{Tq}^{(p)}$ and $f_{TG}^{(p)}$ are related to hadron matrix elements, $\langle p |\bar{q}{q} | p \rangle$ and $\langle p | G^a_{\mu\nu} G^{a\mu\nu} | p \rangle$, with $G^a_{\mu\nu}$ being the field strength tensor of gluon, which are evaluated by lattice simulations with the use of the trace-anomaly relation.

\begin{figure}[t]
\begin{center}
\includegraphics[scale=0.6]{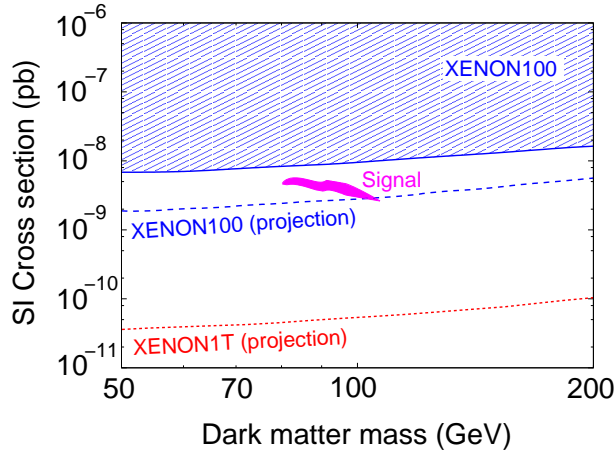}
\caption{\small Scattering cross section between the vector WIMP and a proton as a function of $m_{\rm DM}$. Higgs mass is chosen so that it satisfies WMAP and LHC bounds (95\% C.L.). Current bound and future sensitivity from the XENON100 experiment are also shown.}
\label{fig: direct}
\end{center}
\end{figure}

Result of the scattering cross section is shown in Fig.\ref{fig:
direct}, where {\tt micrOMEGAs} was again used to calculate the cross
section. According to the recent result of the lattice
simulation~\cite{Takeda:2010cw}, the $\pi$-nucleon sigma term has been
set to be $\sigma_{\pi N}= 55$ MeV with $y$-parameter representing the
contribution from the strange content being zero,\footnote{These
parameter choice is also consistent with recent analysis based on chiral
perturbation theory \cite{Alarcon:2011zs}.} which gives
conservative results of dark matter signals. Current bound and future
sensitivity from the XENON100 experiment~\cite{XENON100} are also shown
in the figure. It can be seen that the signal of the vector WIMP is
below the current bound but above the future sensitivity, so that the
signal will be detected in the near future.

\subsection{Indirect detections}

We finally consider the signal of the vector WIMP at indirect detection experiments of dark matter. The annihilation of the vector WIMP would lead to excesses in various cosmic ray spectra. As mentioned in previous subsections, the vector WIMP has the mass of about 100 GeV and annihilates mainly into $W$ and $Z$ boson pairs. Subsequent decays of these $W$ and $Z$ bosons produce anti-protons and gamma-rays, and these contributions are proportional to the annihilation cross section of the dark matter. Stringent constraints or exciting signals would be therefore obtained from indirect detection experiments observing anti-protons and gamma-rays.

\begin{figure}[t]
\begin{center}
\includegraphics[scale=0.6]{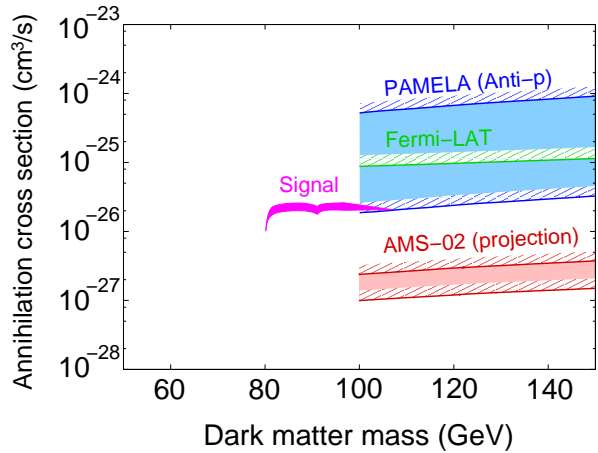}
\caption{\small Constraints on the annihilation cross section of the vector WIMP dark matter as a function of $m_{\rm DM}$, which are obtained from Fermi-LAT and PAMELA experiments. Future sensitivity on the cross section from the AMS-02 experiment is also shown.}
\label{fig: indirect}
\end{center}
\end{figure}

The annihilation cross section predicted by the vector WIMP dark matter after imposing both WMAP and LHC constraints is shown as a magenta-shaded region in Fig.\ref{fig: indirect}. On the other hand, the constraint obtained by the Fermi-LAT experiment, which is observing gamma-rays from milky-way satellites~\cite{Ackermann:2011wa}, is depicted as a green solid line. Another constraint, which is obtained by the PAMELA experiment observing the anti-proton ($\bar{p}$) flux in the cosmic-ray~\cite{Adriani:2010rc}, is also shown in the figure as a blue-shaded region. Since the $\bar{p}$ flux depends on how $\bar{p}$ propagates under the complicated magnetic field of our galaxy and which dark matter profiles we adopt~\cite{Evoli:2011id}, the constraint has large uncertainties as can be seen in the figure. On the contrary, the observation of the $\bar{p}$ flux is very hopeful in the near future. This is because the AMS-02 experiment, which has already been started~\cite{AMS-02}, has better sensitivity than the PAMELA experiment and it is also expected that astrophysical uncertainties related to the $\bar{p}$ propagation are reduced. The future sensitivity to detect the vector WIMP in this experiment is depicted as a red-shaded region with assuming an appropriate propagation model~\cite{Evoli:2011id}. Since the vector WIMP predicts the cross section much above the sensitivity, it will be detected in the near future.
%
% Added for comment on less than 100 GeV region.
%

%Approximate constraints on the dark matter annihilation cross section
%for $m_W < m_{\rm DM} < 100$ GeV are obtained by simply extrapolating the
%experimental upper bounds in Fig.\ref{fig: indirect}
%~\cite{Cholis:2010xb}.
We obtain approximate constraints on the dark matter annihilation cross section
for $m_W < m_{\rm DM} < 100$ GeV by simply extrapolating the
experimental upper bounds in Fig.\ref{fig: indirect}
~\cite{Cholis:2010xb}.  However, due to considerable uncertainties
resulting from {\it e.g.} fragmentation functions, we do not explicitly
show experimental constraints for WIMP mass values below 100 GeV.

%%%%%%%%%%%%%%%%%%%%%%
%%%%% Conclusion %%%%%
%%%%%%%%%%%%%%%%%%%%%%
\section{Conclusions and Discussion}
\label{sec: conclusion}

In conclusion, the higgs mass is shown to be generically in the range
120--125 GeV in scenarios where a neutral vector WIMP that is a
partner of the standard model hypercharge gauge boson accounts for the
observed dark matter abundance.  We call the coincidence between the
predicted values of the higgs mass and the excesses of events observed
at the LHC the ``vector WIMP miracle''.  It should be emphasized that
the interactions between the higgs boson and vector WIMP are
controlled by the U(1)$_Y$ gauge coupling constant and higgs VEV in
the effective low-energy theory, and therefore observables related to
the vector WIMP depend solely on the higgs and vector WIMP masses.
For illustration, we have considered a simple non-linear sigma model
in which the vector WIMP is stabilized by the Z$_2$ parity
corresponding to the exchange of two U(1) sectors.  After imposing the
WMAP and LHC constraints on the model, the vector WIMP mass is
narrowed down to 80--107 GeV.  In this mass region, the WIMP relic
abundance shows a mild dependence on the WIMP mass as the $W$ and $Z$
productions processes compensate the effect of the WIMP mass increase.
The spin-independent WIMP-nucleon cross section is shown to be
$10^{-9}$ pb even in a conservative case; the XENON100 collaboration
will cover the signal region in the near future.  As for the
annihilation cross section of the vector WIMP, the predicted values
are ${\cal O} (10^{-26})~{\rm cm}^3/{\rm s}$; the AMS-02 experiment is
capable of detecting the debris of WIMP annihilations irrespective of
the details of the propagation of anti-protons in our galaxy.

Several comments are in order. 
%We have implicitly assumed that the
%kinetic mixing between the two U(1) gauge bosons is absent in our
%non-linear sigma model.  If the two U(1) gauge bosons mix, the mixing
%parameter is transmitted to the coupling between the higgs boson and
%vector WIMP, leading to a change in the WIMP relic abundance and other
%observables.  However, in UV completion models in which the two U(1)
%sectors are separated at the cutoff scale, such a mixing is
%loop-suppressed and hence negligible at lower energies.
In generic, the $Z_2$ symmetry allows the field strengths of the two
U(1) gauge bosons mix without affecting the stability of the WIMP dark
matter.  Nevertheless, we have not taken into account the kinetic
mixing in our computations as we optimistically expect that the
kinetic mixing vanishes at the cutoff scale from some UV-completion.
If this is the case, the kinetic mixing is induced solely by quantum
corrections, and therefore estimated as
$g_1^2 /(4 \pi)^2 \ln(\Lambda/M_Z) \sim {\cal O}(0.01)$
at the weak scale.  After appropriate redefinitions of the gauge
fields and gauge coupling constants, the kinetic mixing effect turns
out to appear as the scale factor for the dark matter interactions.
Hence, changes in our results caused by the kinetic mixing are also of
the order of ${\cal O}(0.01)$, like other loop corrections.  Since
precise computations at the loop level are beyond our current scope,
we have included neither the kinetic mixing nor other loop-induced
effects.

There are exceptional cases where the simple standard computation of
the WIMP abundance is not valid.  If there exist degenerate particles
that share quantum numbers with the vector WIMP, coannihilations with
such particles alter the WIMP abundance.  If vector WIMP
(co)annihilations take place near a pole, the final WIMP abundance is
significantly decreased.  Both phenomena are crucial in the minimal
UED model, where coannihilations of the 1st Kaluza-Klein (KK)
hypercharge $B$ boson with other 1st KK particles as well as 2nd KK
resonance processes must be taken into account in order to obtain the
correct relic abundance \cite{UEDDM}.

It might be possible to contrive a linear sigma model that has the
same feature as our non-linear sigma model presented.  However, it is
not trivial whether phenomenological constraints are satisfied as a
complicated flavor structure is required in a realistic linear sigma
model.  This issue may be discussed elsewhere.

\section*{Acknowledgments}

This work is supported by Grant-in-Aid for Scientific research from the Ministry of Education, Science, Sports, and Culture (MEXT), Japan, (Nos.\ 22244021 \& 23740169 for S.M.), World Premier International Research Center Initiative (WPI Initiative), MEXT, Japan (S.M.), and the scientific research grants from Hokkai-Gakuen (O.S). M.K., S.M., and O.S. also thank the Yukawa Institute for Theoretical Physics at Kyoto University (YITP). Discussions during the YITP workshop ``Summer Institute 2011'' were useful to develop this work. T.A. and O.S. thank IPMU for their hospitality during the completion of the work.

\end{document}